\documentclass[prl,twocolumn,preprintnumbers,footinbib,tightenlines,superscriptaddress]{revtex4}

\pdfoutput=1
\usepackage[english]{babel}
\usepackage{amsmath,amssymb,amsfonts, bm,bbm,slashed, subdepth}
\usepackage{graphicx}
\usepackage[sort&compress]{natbib}
\usepackage{xcolor}
\usepackage[normalem]{ulem}
\usepackage{hyperref}
\usepackage{cleveref}
\definecolor{red}{rgb}{1.0, 0, 0}
\usepackage{enumerate}
\usepackage{epsfig, subfigure}
\usepackage{setspace}
\usepackage{booktabs, tabularx}
\usepackage{units}
\usepackage{placeins}
\usepackage{multirow}
\usepackage{mathtools}

\newcommand{\beq}{\begin{equation}}
\newcommand{\eeq}{\end{equation}}

\newcommand{\Mp}{M_{\scriptscriptstyle\rm Pl}}
\newcommand{\calO}{\mathcal{O}}
\newcommand{\calN}{\mathcal{N}}

\makeatletter
\newcommand{\vast}{\bBigg@{3.5}}
\newcommand{\Vast}{\bBigg@{4.5}}
\makeatother

\begin{document}

\preprint{IPMU18-0178}
\title{What does Inflation say about Dark Energy given the Swampland Conjectures?}

\author{Chien-I Chiang}
\email{chienichiang@berkeley.edu}
\affiliation{Department of Physics, University of California, Berkeley, CA 94720, USA}
\affiliation{Ernest Orlando Lawrence Berkeley National Laboratory, Berkeley, CA 94720, USA}

\author{Jacob M. Leedom}
\email{leedoj@berkeley.edu}
\affiliation{Department of Physics, University of California, Berkeley, CA 94720, USA}
\affiliation{Ernest Orlando Lawrence Berkeley National Laboratory, Berkeley, CA 94720, USA}

\author{Hitoshi Murayama}
\email{hitoshi@berkeley.edu, hitoshi.murayama@ipmu.jp}
\affiliation{Department of Physics, University of California, Berkeley, CA 94720, USA}
\affiliation{Kavli Institute for the Physics and Mathematics of the
  Universe (WPI), University of Tokyo,
  Kashiwa 277-8583, Japan}
\affiliation{Ernest Orlando Lawrence Berkeley National Laboratory, Berkeley, CA 94720, USA}

\begin{abstract}
We discuss the relations between swampland conjectures and observational constraints on both inflation and dark energy.  Using the  requirement $|\nabla V|\geq c V$, with $c$ as a {\it universal}\/ constant whose value can be derived from inflation, there may be no observable distinction between constant and non-constant models of dark energy.  However, the latest modification of the above conjecture, which utilizes the second derivative of the potential, opens up the opportunity for observations to determine if the dark energy equation of state deviates from that of a cosmological constant.  We also comment on the observability of tensor fluctuations despite the conjecture that field excursions are smaller than the Planck scale.
\end{abstract}

\maketitle

\subsection{Introduction}\vspace{-10pt}
The discovery of the accelerating expansion of the Universe \cite{Perlmutter:1998np,Riess:1998cb} was a huge surprise to the community.  Because gravity only {\it pulls}\/, it should put a brake on the expansion of the Universe after the Big Bang and hence the expansion should decelerate.  Acceleration implies there is a substance in the Universe that {\it pushes}\/ the expansion.  It was dubbed {\it dark energy}\/.
The most discussed candidate for dark energy is the cosmological constant $\Lambda$, a finite energy density of the vacuum, due to the simple way it can be implemented into cosmological models based on general relativity.  However, despite being consistent with data \cite{Aghanim:2018eyx}, the 120 orders of magnitude difference between the observed vacuum energy density ($\rho \approx ({\rm meV})^{4}$) and the na\"ive theoretical expectation ($\rho \approx \Mp^{4}$) still remains the most challenging problem in  modern physics \cite{Weinberg:1988cp}.

Since dark energy and the cosmological constant problem inevitably involve quantum gravity, string theory, as a theory of quantum gravity, should address these topics. The attempts to construct de Sitter solutions (spacetime solutions to general relativity with a positive $\Lambda$) in string theory \cite{Bousso:2000xa,Giddings:2001yu,Kachru:2003aw} have lead to the notion of the string landscape. The landscape consists of an enormous number of vacua, each described by different low-energy effective field theories (EFTs) of different fields and parameters. String theory therefore supports the anthropic argument \cite{Weinberg:1987dv}, namely that the value of the observed dark energy density is what it is because otherwise human civilization could not exist. If we really live in a (meta-)stable vacuum in the string landscape where a constant vacuum energy explains dark energy, then there is no point in measuring the dark energy equation of state parameter $w=p/\rho$, where $p$ and $\rho$ are the pressure and energy density of the dark energy, respectively.

String theory seems to lead to many possible low-energy EFTs, so conversely one can ask what criteria a given low-energy EFT should satisfy in order to be contained in the string landscape. For the last decade, several criteria of this kind, dubbed \textit{swampland conjectures}, have been proposed \cite{Vafa:2005ui,Ooguri:2006in,ArkaniHamed:2006dz}. These can have important cosmological implications. For instance, one of the relatively well-established conjectures is the \textit{distance swampland conjecture} \cite{Ooguri:2006in,Palti:2015xra,Baume:2016psm,Klaewer:2016kiy,Valenzuela:2016yny,Blumenhagen:2017cxt,Palti:2017elp,Lust:2017wrl,Hebecker:2017lxm,Cicoli:2018tcq,Grimm:2018ohb,Heidenreich:2018kpg,Blumenhagen:2018nts,Lee:2018urn} which implies that scalar fields in a low-energy EFT of a consistent theory of quantum gravity cannot have field excursions much larger than the Planck scale since otherwise an infinite tower of states becomes exponentially light and the validity of the EFT breaks down. In other words, one has the constraint
\beq
\Delta \phi \lesssim \alpha \Mp, \qquad \alpha \approx O(1).
\label{eq:distance}
\eeq

In the context of inflation, field excursions are related to the tensor-to-scalar ratio $r$ by the Lyth bound \cite{Copeland:1994vg},
\beq
\frac{\Delta \phi}{\Mp} \simeq \sqrt{\frac{r}{8}}\, \calN
\label{eq:Lyth}
\eeq
where $ \calN$ is the number of $e$-folds of inflationary expansion.  Clearly the distance conjecture, Eq.~\eqref{eq:distance}, limits the possibility of measuring tensor modes and hence primordial B-modes in the cosmic microwave background (CMB).  Naively, with ${\cal N}\gtrsim 50$, we find $r \lesssim 0.003$, which is on the edge of observability for future experiments \cite{Abazajian:2016yjj,Suzuki:2018cuy}.

The attempts to construct de Sitter solutions or inflationary models in string theory \cite{Kachru:2003aw,Kachru:2003sx,Silverstein:2003hf,Balasubramanian:2005zx,Baumann:2006th,Westphal:2006tn,Baumann:2007ah,Dong:2010pm,Rummel:2011cd,Blaback:2013fca,Cicoli:2013cha,Cicoli:2015ylx} have sparked discussions on various issues with such constructions, as well as no-go theorems
\cite{Maldacena:2000mw,Townsend:2003qv,Hertzberg:2007wc,Covi:2008ea,McGuirk:2009xx,Caviezel:2008tf,Caviezel:2009tu,deCarlos:2009fq,Wrase:2010ew,Shiu:2011zt,Green:2011cn,Gautason:2012tb,Bena:2012vz,Blaback:2012nf,Bena:2014jaa,Danielsson:2014yga,Kutasov:2015eba,Quigley:2015jia,Dasgupta:2014pma,Junghans:2016abx,Junghans:2016uvg,Andriot:2016xvq,Moritz:2017xto,Sethi:2017phn,Andriot:2017jhf,Danielsson:2018ztv}.
 Motivated by the obstructions encountered in various attempts, the \textit{de Sitter swampland conjecture} was proposed \cite{Obied:2018sgi}, which states that the scalar potential of a low-energy limit of quantum gravity must satisfy
\begin{equation}
	\Mp \lvert\nabla V\rvert \ge c \,V, \qquad c \approx O(1) > 0 \label{dS}
\end{equation}
where $\nabla$ denotes the gradient with respect to the field space, and the norm of the gradient is defined by the metric on field space.  Whether the conjecture holds true is still an open debate \cite{Kallosh:2014wsa,Ferrara:2014kva,Bergshoeff:2015jxa,Bergshoeff:2015tra,Hasegawa:2015bza,Kallosh:2015nia,Polchinski:2015bea,Kallosh:2016aep,Aalsma:2018pll,Cicoli:2018kdo,Dasgupta:2018rtp,Roupec:2018mbn,Andriot:2018ept,Andriot:2018wzk,Conlon:2018eyr,Akrami:2018ylq,Kallosh:2018wme,Kachru:2018aqn,Moritz:2018ani,Bena:2018fqc,Gautason:2018gln}. Yet, even before the debate is settled, it is interesting and important to investigate both its consequences in cosmology and potential modifications or extensions \cite{Choi:2018rze,Heisenberg:2018rdu,Heisenberg:2018yae,Marsh:2018kub,Murayama:2018lie,Kinney:2018nny,Damian:2018tlf,Ben-Dayan:2018mhe,Matsui:2018bsy,Colgain:2018wgk,Denef:2018etk,Dias:2018ngv,Kehagias:2018uem,Garg:2018reu,Achucarro:2018vey,Das:2018hqy,Wang:2018duq,Brandenberger:2018wbg,Han:2018yrk,Brandenberger:2018xnf,Dimopoulos:2018upl,Ellis:2018xdr,Lin:2018kjm,Hamaguchi:2018vtv,Kawasaki:2018daf,Motaharfar:2018zyb,Ashoorioon:2018sqb,Das:2018rpg,Wang:2018kly,Fukuda:2018haz,Hebecker:2018vxz,Olguin-Tejo:2018pfq,Garg:2018zdg,Park:2018fuj,Blaback:2018hdo,Schimmrigk:2018gch,Lin:2018rnx}. The primary implication of this condition is that the observed positive energy density of our Universe should correspond to the potential of a rolling quintessence field rather than a positive $\Lambda$ \cite{Agrawal:2018own}. The fact that one can easily embed any quintessence model into supergravity \cite{Brax:2009kd,Chiang:2018jdg} in a rather simple fashion, despite the difficulty that supersymmetry breaking generically spoils the flatness of the quintessence potential,  is also encouraging. This raises the hope that $w\neq -1$ might be detected.

The de Sitter conjecture forbids (meta-)stable vacua with positive energy density, so it is not surprising that the inflationary paradigm has apparent conflicts with the conjecture and one may call for a paradigm shift. Nonetheless, one can also adopt a conservative approach and regard the conjecture as a parametric constraint where the inequality holds but the number $c$ may not be strictly $\calO(1)$ \cite{Dias:2018ngv}. From this perspective, constraints on inflation can then be used to constrain $c$.

However, if we follow this route, the optimism that one can observe $w\neq -1$ is greatly diminished. To see this, recall that in single-field slow-roll inflation, the slow-roll parameters of the potential are defined as
\beq
\epsilon_V \equiv \frac{\Mp^2}{2}\left(\frac{V'}{V} \right)^2, \quad
\eta_V \equiv \Mp^2 \frac{V''}{V}\,,
\eeq
where the primes denote derivatives with respect to the inflaton. The distance conjecture limits the inflaton field excursion $\Delta \phi \approx \sqrt{2\epsilon_V}\, {\cal N} \lesssim \calO(1)$ and therefore the necessary number of $e$-folds ${\cal N}\approx 50 $ forces $c \lesssim \sqrt{2\epsilon_V}  \lesssim {\cal N}^{-1} \sim 0.02$. On the other hand, the number $c$ in Eq.~\eqref{dS} is meant to be \textit{universal} in a given EFT.  Therefore, the current accelerating expansion must involve a quintessence field $Q$ whose potential $V_{Q}$ must satisfy
\beq
\label{eq:deb1}
	1+w = \frac{2(V_Q^\prime)^2}{(V_Q^\prime)^2 + 6V_Q^2} > \frac{2c^2}{6+c^2} \equiv \Delta
	\gtrsim 1.33\times10^{-4}.
\eeq
Although this does not exclude observable quintessence, given the fact that so far almost all observations are consistent with a cosmological constant, such a small lower bound on possible deviation of $w$ from $-1$ makes it questionable if it is worthwhile to push the sensitivity of the observations further. We may never know whether the Universe is de Sitter or quintessence.

However, the original de Sitter conjecture, Eq.~(\ref{dS}), was so strong that even the Higgs potential was in tension with it \cite{Denef:2018etk}. The conjecture was also in tension with the well-understood supersymmetric AdS solutions \cite{Conlon:2018eyr}. Recently the \textit{refined de Sitter swampland conjecture} was proposed \cite{Garg:2018reu,Ooguri:2018wrx}, which states that the scalar potential of a low-energy theory that can be consistently coupled to quantum gravity should satisfy \textit{either}
\begin{eqnarray}
	 &\Mp \lvert\nabla V\rvert \ge c \,V , \qquad c \approx O(1)>0, & \label{dS1} \\
&\textit{or}& \nonumber \\
&\Mp^{2} \text{min}(\nabla_i\nabla_j V) \le - c^\prime V ,  \qquad c' \approx O(1)>0, &\label{dS2}
\end{eqnarray}
where min(...) denotes the minimum eigenvalue of the Hessian $\nabla_i\nabla_j V$ in an orthonormal frame of the scalar field space.  With this refinement, the aforementioned conflicts with the Higgs potential and the SUSY AdS solutions are resolved. The refined conjecture also raises new possibilities for inflation. In particular, one can evade the strict bound on $c$ arising from the distance conjecture by having the scalar potential satisfy the second condition Eq.~(\ref{dS2}) of the new conjecture during part (or all) of inflation. As such, one may regain the hope that observable time-varying dark energy with $w\neq -1$ can be obtained. See also \cite{Agrawal:2018rcg} for a recent discussion on $w$ in consideration of the refined dS conjecture. \vspace{-20pt}

\subsection{Single-Field Slow-Roll Inflation Models}\vspace{-10pt} 
Due to the above tension between the de Sitter conjecture and the requirements of inflation, we assume that the inflaton potential switches from one de Sitter condition to another as the inflaton rolls, an idea also utilized in \cite{Fukuda:2018haz}. To be specific, we take the following step-function approach to keep the discussion general and simple: we apply the first condition, Eq.~\eqref{dS1}, for the initial ${\cal N}_{1}$ e-folds and apply the second condition, Eq.~\eqref{dS2}, for the remaining ${\cal N}_{2} = {\cal N}_{tot} - {\cal N}_1$ $e$-folds.  In our analysis we set $\calN_{tot} =50$. We assume $\epsilon_{V}$ and $\eta_{V}$ are approximately constant for each interval so that we have
\begin{equation}
\label{pw1}
		\sqrt{2\epsilon^{(1)}_V} \ge c \text{ and } \eta^{(2)}_V \le -c^\prime .
\end{equation}
Additionally, Eq.~\eqref{eq:distance} requires that 
\begin{equation}
\label{pw2}
	\sqrt{2\epsilon^{(1)}_{V}} {\cal N}_{1} + \sqrt{2\epsilon^{(2)}_V} {\cal N}_{2} \leq \alpha \sim O(1).
\end{equation}
To maximize $c$, we assume $\epsilon^{(2)}_V < 10^{-4}$ so that the contribution of the second era to Eq.~\eqref{eq:distance} is negligible. Combining Eq.~\eqref{pw1} and Eq.~\eqref{pw2}, we have
\begin{equation}
	c  < \frac{\alpha - \sqrt{2\epsilon^{(2)}_V}{\cal  N}_2}{{\cal N}_1}\ .
\end{equation}

We can also obtain a bound for $c^\prime$ from the spectral tilt $n_s= 1- 2\epsilon -\eta$, where the Hubble slow-roll parameters are  
\begin{equation}
\epsilon = -\frac{\dot{H}}{H^{2}}, \qquad \eta = \frac{\dot{\epsilon}}{H\epsilon}\ .
\end{equation}
For single-field inflation models, these are related to the slow-roll parameters of the potential as $\epsilon_V = \epsilon$ and $\eta_{V} = 2\epsilon - \frac{1}{2}\eta$.  Therefore, we can constrain $\eta_{V}$ and hence the second parameter of the  refined de Sitter conjecture as

\begin{align}
	c^\prime < \frac{1}{2} \bigg(1-n_s(k)-6\epsilon^{(2)}_V\bigg)\ ,
\end{align}
where we are allowing for a $k$-dependent spectral tilt. Since we assume $\epsilon^{(2)}_V$ is small, our bounds simplify to
\begin{align}
\label{eq:bounds}
		(c^\prime, c) < \left(\frac{1-n_s(k)}{2} , \frac{\alpha}{{\cal N}_1} \right) . 
\end{align}
Eq.~\eqref{eq:bounds} is valid until ${\cal N}_1 = {\cal N}_{tot}$, at which point the derivation on the bound of $c^\prime$ above no longer applies, and the only constraint one finds is that $c < \alpha/\calN_{tot}$. To proceed, we utilize the Planck analysis based on TT, TE, EE, lowE, lensing and BAO \cite{Aghanim:2018eyx}, which gives

\begin{align}
	dn_{s}/d \ln k &= -0.0041 \pm 0.0067, \\
	n_{s} &= 0.9659 \pm 0.0040,
\end{align}
at $k_{*}=0.05$Mpc$^{-1}$.
We add errors in quadrature, ignoring correlations, and use

\begin{align}
	n_{s}(k) = 0.9659 &- 0.0041 \ln \frac{k}{k_{*}} \nonumber \\
	& \pm \sqrt{(0.0040)^{2} + \left(0.0067 \ln \frac{k}{k_{*}}\right)^{2} }\ . \label{nsk}
\end{align}

\begin{figure}[tb]
 \begin{center}
  \includegraphics[width=1.2\linewidth]{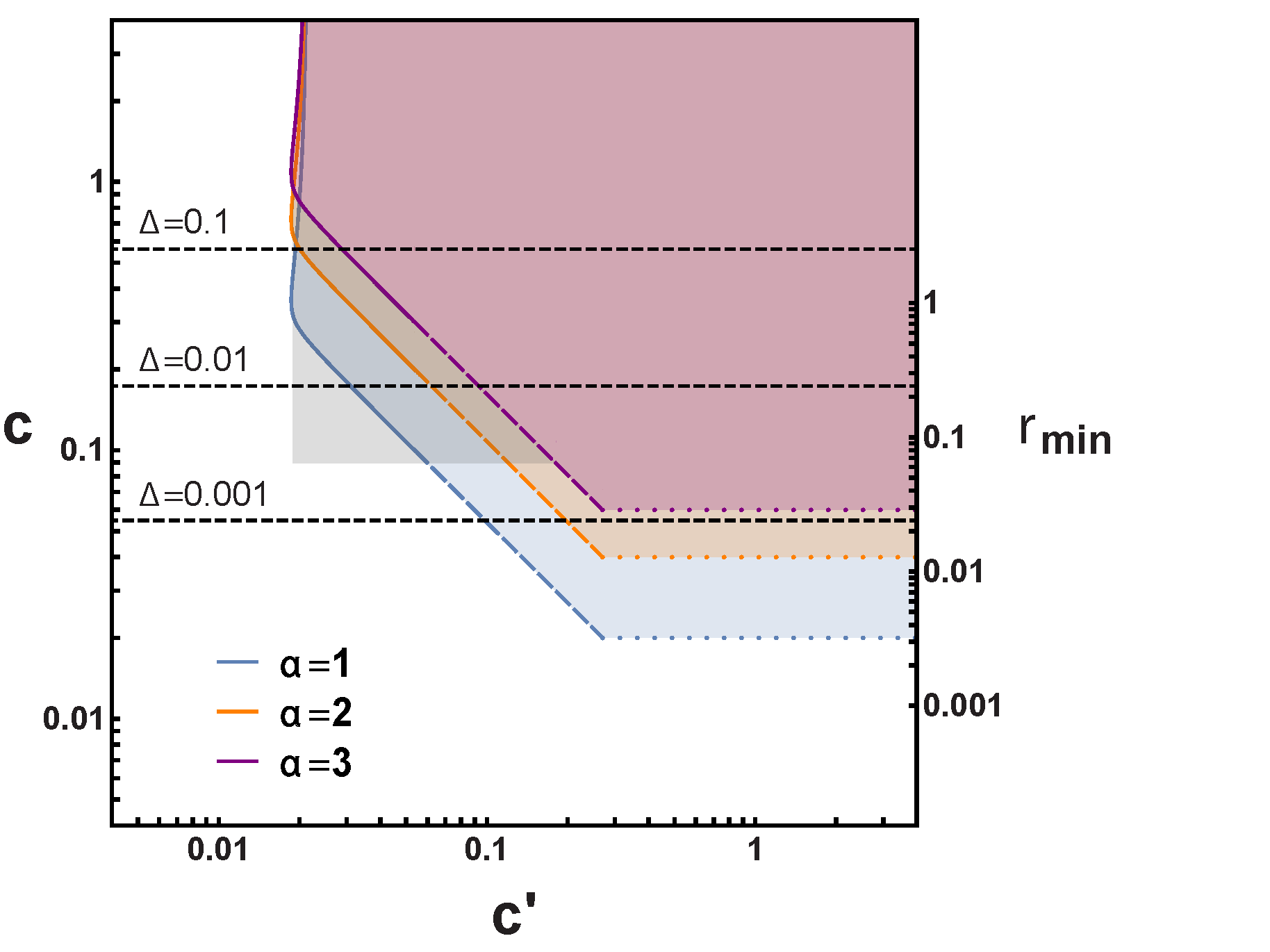}
 \end{center}
\caption{\small\sl 
Bounds on swampland parameters for generic single-field inflation models at the $1\sigma$ level assuming the running of $n_s$ can be extended to $\calN_{tot} = 50$ $e$-folds. The unshaded region is the allowed parameter space.  The solid lines are for $\calN_1 \leq 10$; the dashed lines are for $10<\calN_1<50$, and the horizontal dotted lines correspond to $\calN_1=50$, \text{i.e.} the first constraint Eq.~\eqref{dS1} applies to the whole inflationary period. The values of $c$ excluded by \cite{Akrami:2018odb} are shaded in grey.   We required the distance conjecture with $\Delta \phi \leq \alpha M_{Pl}$, and display the minimum values for $ 1+w \geq \Delta $ with black dashed lines. With the original de Sitter conjecture, $c$ had to be below the dotted horizontal lines but there were no constraints on $c'$.}
\label{fig:swpara1}
\end{figure}

A smaller $n_{s}$ allows for larger $c'$ in Eq.~\eqref{eq:bounds}, so we take the $1\sigma$ allowed lower end in order to place our bounds. The weak correlation between $n_{s}$ and $d n_{s}/d \ln k$ we see in Fig.~26 of~\cite{Aghanim:2018eyx} actually works in our favor and ignoring correlation is therefore the more conservative approach ({\it i.e.}\/, gives a smaller allowed range) \footnote{The Planck 2018 paper \cite{Aghanim:2018eyx} also shows the analysis where they allow for the running of running $d^{2}n_{s}/d \ln k^{2}$.  Unfortunately they do not show the correlation and we cannot use it for our purposes.  In fact, the extrapolation of $n_{s}(k)$ to small scales from the Planck data is most likely too restrictive, as the allowed range for the primordial power $P_{\zeta}(k)$ blows up for $k\gtrsim 0.2~{\rm Mpc}^{-1}$ (see Fig.~20 in \cite{Akrami:2018odb}).}. Using the simple relationship $\calN_1 = \ln \left(k/a_0H_0\right)$, where $a_0$ is the present scale factor and $H_0$ is the present Hubble scale, we can constrain the swampland parameters  in single-field inflation as shown in Fig.~\ref{fig:swpara1}. The current CMB constraints on the spectral index and its running are limited to $\calN_1\lesssim 10$. This range is denoted by the  solid lines in Fig.~\ref{fig:swpara1}. Beyond this there are no strong observational constraints and we extend our analysis by extrapolating Eq.~(\ref{nsk}) to $\calN_1 \geq 10$ shown by the dashed lines in Fig.~\ref{fig:swpara1}. The unshaded regions indicate values of $(c^\prime, c)$ that satisfy the above inequalities. The vertical asymptotes correspond to satisfying Eq.~\eqref{dS2} for the entirety of the inflationary epoch, $\calN_1=0$, so that $c$ is left completely arbitrary but $c^\prime$ has a strict upper bound that is much less than the $\mathcal{O}(1)$ expectation. The horizontal dotted lines correspond to satisfying the first constraint Eq.~\eqref{dS1} for all of inflation, $\calN_2=0$, which leaves $c^\prime$ arbitrary but severely limits $c$. The horizontal black dashed lines indicate the lowest values of $c$ that yield the given $\Delta$ defined in Eq.~\eqref{eq:deb1} as the lower bound on $1+w$ from the constraint Eq.~\eqref{dS1}. Finally, the grey region excludes values of $c$ that may satisfy Eq.\eqref{eq:bounds}, depending on the value of $\alpha$, but conflicts with the constraint $r_{0.002} < 0.064$ \cite{Akrami:2018odb}, as $r=16\epsilon \geq 8c^2$.  The grey excluded region has a left vertical boundary since the constraint applies only to $k>$ 0.002 Mpc$^{-1}$.

	We also comment on the observability of the tensor mode $r$.  The swampland distance conjecture, Eq.~(\ref{eq:distance}), combined with the Lyth bound, Eq.~(\ref{eq:Lyth}), is normally believed to disfavor observably large $r$, assuming $\alpha \approx 1$.  The best sensitivity anticipated in the  future is $r \sim 10^{-3}$ \cite{Abazajian:2016yjj,Suzuki:2018cuy}.  There is a parameter region in Figure \ref{fig:swpara1} where $r\geq r_{\rm min}\equiv8c^2$ is close to the current observational bound. Physically this is because, in our spirit of a step function approximation, we can allow for a brief initial period, say ${\cal N}_{0} \sim 4$, where the upper bound on $\epsilon$ from the distance conjecture, $ \sl   \epsilon \lesssim {\cal N}_{0}^{-2}/2 \sim 0.03$, is relaxed. Thus it is possible to have $r$ large enough to saturate the observational bound at low $\ell$.  This is encouraging, especially for space-born CMB $B$-mode experiments such as LiteBIRD \cite{Suzuki:2018cuy}.

\subsection{Multi-Field Slow-Roll Inflation Models}\vspace{-10pt} 
The constraints discussed above are due to the tight relations between $n_{s}$, $\epsilon_{V}$, $\eta_{V}$, and $r$ in single-field slow-roll inflation models.  
It is natural to ask whether the constraints can be relaxed in multi-field models. In our analysis below, we take the conservative assumption that the swampland distance conjecture applies to the proper length of the trajectory, instead of the geodesic distance between the starting and ending points in the field space. 

\begin{figure}[tb]
 \begin{center}
  \includegraphics[width=1.1\linewidth]{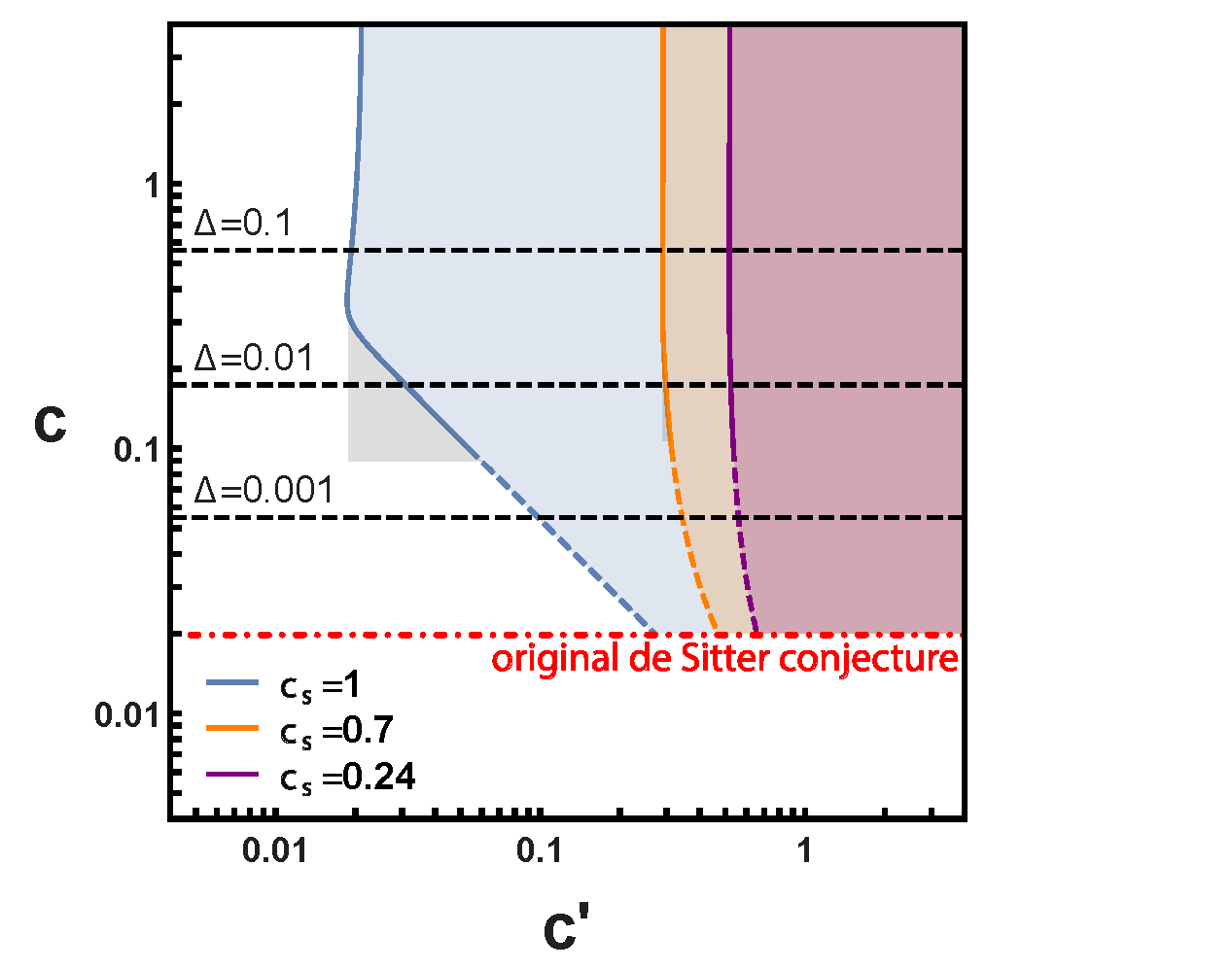}
 \end{center}
\caption{\small\sl 
Bounds on swampland parameters for generic multi-field inflation models. We took $\alpha=1$ and $M=H$. $c_{s}$ is the sound speed for fluctuations, and the rest is the same as in Fig.~\ref{fig:swpara1}. With the original de Sitter conjecture, Eq.~\eqref{dS},  and single-field slow-roll models, $c$ had to be below the red dot-dashed horizontal line.}
\label{fig:swpara2}
\end{figure}

We discuss here a class of multi-field models where directions orthogonal to the slow-roll direction are massive, $M \gtrsim H$.  The inflaton therefore rolls near the bottom of the valley, which has ``bends'' in the multi-dimensional field space.  
The main difference here is that the local angular velocities of the inflaton around the bends can modify the effective sound speed $c_s$ of fluctuations. As a result, we have the modified relation \cite{Hetz:2016ics}
\begin{align}
&12 \eta_V =  (c_s^{-2} - 1 ) \frac{M^2}{H^2}  + 2 \frac{M^2}{H^2}  +  3  (4 \epsilon - \eta) 
\qquad \quad \nonumber \\ 
& -  2  \sqrt{  \left(   \frac{M^2}{H^2} -  \frac{3}{2}  (4 \epsilon - \eta )\right)^2   +  9 (c_s^{-2} - 1 ) \frac{M^2}{H^2}    } \ .  \quad \label{eta_V_two_fields}
\end{align}
Here, $\eta_V$ is the minimum eigenvalue of the Hessian and $M$ is the effective mass of the field orthogonal to the slow-roll direction, and $c_s$ is given by
\begin{equation}
	c_{s}^{-2} = 1 + \frac{4\Omega^{2}}{M^{2}}\ ,
\end{equation}
where $\Omega$ is the local angular velocity describing the bend of the inflaton trajectory in the potential.  Note that in the limit $\Omega \rightarrow 0$, the sound speed reduces to unity and $\eta_{V}$ to the expression of the single-field models.  Allowing for a significant deviation of $c_{s}$ from unity relaxes the constraints on $(c, c')$, as shown in Fig.~\ref{fig:swpara2}, where we set $M=H$. This allows for larger values of $c$ and $c'$ compared to the single-field case, which are preferred by the swampland conjecture. Note that lowering the sound speed further will not achieve $\mathcal{O}(1)$ values for $c^\prime$ because our scenario relies on having negative $\eta_V$. As $c_s$ is reduced from unity, $\eta_V$ initially becomes more negative and widens the allowed parameter space. Beyond some critical value $c_{s}\approx 0.3$, further reduction of $c_s$ makes $\eta_V$ less negative, thereby narrowing the allowed parameter space. For $c_{s} \lesssim 0.2$, $\eta_V$ becomes positive and our analysis no longer holds. Empirically, we find that $c_s\sim0.24$ maximizes the allowed parameter region in the $(c', c)$-plane. The grey shaded regions again correspond to experimental constraints on $r=16\epsilon c_s$, but their area is greatly reduced as $c_s$ decreases.

It is also interesting to note that we expect primordial equilateral and orthogonal non-Gaussianities once $c_{s} \neq 1$ in this class of models \cite{Hetz:2016ics},
\begin{align}
f_{\rm NL}^{\rm equil} &= - (c_s^{-2} - 1) ( 0.275 + 0.078 c_s^2 ), \label{fnl-1}\\
f_{\rm NL}^{\rm ortho} &=  (c_s^{-2} - 1) ( 0.0159 - 0.0167 c_s^2 ). \label{fnl-2}
\end{align}
Here we have ignored the third order parameter. The current observational constraint on the sound speed is $c_s \geq0.024$ (see Eq.~(89) of \cite{Ade:2015ava}),
which is an order of magnitude below the limit we can reach in our setup, as shown in Fig.~\ref{fig:swpara2}. Future observations combining CMB lensing, galaxy and 21cm surveys, Lyman $\alpha$ forest, {\it etc.}\/ have the potential to improve the constraint on $f_{\rm NL}$ by an order of magnitude or more \cite{CMBS4test}. 

\vspace{-10pt}

\subsection{Implications for Dark Energy}\vspace{-10pt}
The de Sitter conjecture states that constants $c$ and $c^\prime$ are {\it universal}\/ and should apply to all sectors in a given EFT. Therefore, we can use inflationary physics to get a handle on the values of $c$ and $c^\prime$ and apply this knowledge to the quintessence potential $V_{Q}$. 
When this argument is applied to single-field inflation models with conjectures Eq.~\eqref{dS} and Eq.~\eqref{eq:distance}, one deduces that there may be little hope in finding $w\neq -1$ due to the small lower bound seen in Eq.~\eqref{eq:deb1}. This depressing outlook is drastically changed in light of  Eqs.~(\ref{dS1}) and (\ref{dS2}), as Fig.~\ref{fig:swpara1} illustrates. We see that the refined de Sitter conjecture has allowed for the possibility of having $\Delta$ bounded from below such that it must be larger than a few per cent and should be observable to experiments. Current and future experiments, such as DES \cite{DES}, HSC \cite{HSC}, DESI \cite{DESI}, PFS \cite{PFS}, LSST \cite{LSST}, Euclid \cite{Euclid}, and WFIRST \cite{WFIRST}, are aiming for an accuracy of about a percent in $w$.  The cost for this is that $c^\prime$ must be much lower than the $\mathcal{O}(1)$ expectation of \cite{Garg:2018reu,Ooguri:2018wrx} in the single-field case. This seems to indicate that single-field inflation falls more in line with the modified de Sitter conjecture discussed in \cite{Murayama:2018lie}, where the smallest Hessian eigenvalue needs only be negative when $\lvert \nabla V\rvert < cV$. 

This state of affairs is altered by considering multi-field inflation models. Not only could $\Delta$ be forced to be as large as several per cent, it is also possible to have both $c$ and $c^\prime$ approximately $\mathcal{O}(1)$ as long as the sound speed is low enough, as seen in  Fig.~\ref{fig:swpara2}.  In either the single-field or multi-field scenario, a better theoretical understanding of the magnitude of $c^\prime$ is essential to understand the consistency of the swampland conjectures and inflation.






\subsection{Conclusions}\vspace{-10pt}

In this Letter, we studied the consequences of the latest swampland conjecture on inflation and dark energy.  The original de Sitter conjecture 
 raised the hope that measuring the dark energy equation of state  $w$ would be promising while simultaneously dashing that hope since consistency with single-field inflation suggests that the deviation from $w=-1$ would likely be unobservable. As we have shown, this situation is much more encouraging with the refined de Sitter conjecture.
Not only could $w\neq -1$ be observable even with a single-field inflationary scenario, but tensor modes could be as well. If one considers multi-field inflationary scenarios, then the prospect for observing $w \neq -1$ is better and one gains improved agreement with the swampland conjectures.



\begin{acknowledgments}
We would like to thank Katelin Schutz for useful discussions and comments on the manuscript.
H.M., CIC, and J.M.L. were supported by the U.S. DOE Contract DE-AC02-05CH11231.
H.M. was also supported by the NSF grant PHY-1638509, by the JSPS Grant-in-Aid for Scientific Research (C) (17K05409), MEXT Grant-in-Aid for Scientific Research on Innovative Areas (15H05887, 15K21733), by WPI, MEXT, Japan, and  by the Binational Science Foundation  (grant No. 2016153).  
\end{acknowledgments}

\bibliographystyle{utphys}
\bibliography{Refs}

\providecommand{\href}[2]{#2}\begingroup\raggedright\begin{thebibliography}{100}

\bibitem{Perlmutter:1998np}
{\bfseries Supernova Cosmology Project} Collaboration, S.~Perlmutter {\em
  et~al.}, ``{Measurements of Omega and Lambda from 42 high redshift
  supernovae},'' \href{http://dx.doi.org/10.1086/307221}{{\em Astrophys. J.}
  {\bfseries 517} (1999) 565--586},
\href{http://arxiv.org/abs/astro-ph/9812133}{{\ttfamily arXiv:astro-ph/9812133
  [astro-ph]}}.

\bibitem{Riess:1998cb}
{\bfseries Supernova Search Team} Collaboration, A.~G. Riess {\em et~al.},
  ``{Observational evidence from supernovae for an accelerating universe and a
  cosmological constant},'' \href{http://dx.doi.org/10.1086/300499}{{\em
  Astron. J.} {\bfseries 116} (1998) 1009--1038},
\href{http://arxiv.org/abs/astro-ph/9805201}{{\ttfamily arXiv:astro-ph/9805201
  [astro-ph]}}.

\bibitem{Aghanim:2018eyx}
{\bfseries Planck} Collaboration, N.~Aghanim {\em et~al.}, ``{Planck 2018
  results. VI. Cosmological parameters},''
\href{http://arxiv.org/abs/1807.06209}{{\ttfamily arXiv:1807.06209
  [astro-ph.CO]}}.

\bibitem{Weinberg:1988cp}
S.~Weinberg, ``{The Cosmological Constant Problem},''
\href{http://dx.doi.org/10.1103/RevModPhys.61.1}{{\em Rev. Mod. Phys.}
  {\bfseries 61} (1989) 1--23}.

\bibitem{Bousso:2000xa}
R.~Bousso and J.~Polchinski, ``{Quantization of four form fluxes and dynamical
  neutralization of the cosmological constant},''
  \href{http://dx.doi.org/10.1088/1126-6708/2000/06/006}{{\em JHEP} {\bfseries
  06} (2000) 006},
\href{http://arxiv.org/abs/hep-th/0004134}{{\ttfamily arXiv:hep-th/0004134
  [hep-th]}}.

\bibitem{Giddings:2001yu}
S.~B. Giddings, S.~Kachru, and J.~Polchinski, ``{Hierarchies from fluxes in
  string compactifications},''
  \href{http://dx.doi.org/10.1103/PhysRevD.66.106006}{{\em Phys. Rev.}
  {\bfseries D66} (2002) 106006},
\href{http://arxiv.org/abs/hep-th/0105097}{{\ttfamily arXiv:hep-th/0105097
  [hep-th]}}.

\bibitem{Kachru:2003aw}
S.~Kachru, R.~Kallosh, A.~D. Linde, and S.~P. Trivedi, ``{De Sitter vacua in
  string theory},'' \href{http://dx.doi.org/10.1103/PhysRevD.68.046005}{{\em
  Phys. Rev.} {\bfseries D68} (2003) 046005},
\href{http://arxiv.org/abs/hep-th/0301240}{{\ttfamily arXiv:hep-th/0301240
  [hep-th]}}.

\bibitem{Weinberg:1987dv}
S.~Weinberg, ``{Anthropic Bound on the Cosmological Constant},''
\href{http://dx.doi.org/10.1103/PhysRevLett.59.2607}{{\em Phys. Rev. Lett.}
  {\bfseries 59} (1987) 2607}.

\bibitem{Vafa:2005ui}
C.~Vafa, ``{The String landscape and the swampland},''
\href{http://arxiv.org/abs/hep-th/0509212}{{\ttfamily arXiv:hep-th/0509212
  [hep-th]}}.

\bibitem{Ooguri:2006in}
H.~Ooguri and C.~Vafa, ``{On the Geometry of the String Landscape and the
  Swampland},'' \href{http://dx.doi.org/10.1016/j.nuclphysb.2006.10.033}{{\em
  Nucl. Phys.} {\bfseries B766} (2007) 21--33},
\href{http://arxiv.org/abs/hep-th/0605264}{{\ttfamily arXiv:hep-th/0605264
  [hep-th]}}.

\bibitem{ArkaniHamed:2006dz}
N.~Arkani-Hamed, L.~Motl, A.~Nicolis, and C.~Vafa, ``{The String landscape,
  black holes and gravity as the weakest force},''
  \href{http://dx.doi.org/10.1088/1126-6708/2007/06/060}{{\em JHEP} {\bfseries
  06} (2007) 060},
\href{http://arxiv.org/abs/hep-th/0601001}{{\ttfamily arXiv:hep-th/0601001
  [hep-th]}}.

\bibitem{Palti:2015xra}
E.~Palti, ``{On Natural Inflation and Moduli Stabilisation in String Theory},''
  \href{http://dx.doi.org/10.1007/JHEP10(2015)188}{{\em JHEP} {\bfseries 10}
  (2015) 188},
\href{http://arxiv.org/abs/1508.00009}{{\ttfamily arXiv:1508.00009 [hep-th]}}.

\bibitem{Baume:2016psm}
F.~Baume and E.~Palti, ``{Backreacted Axion Field Ranges in String Theory},''
  \href{http://dx.doi.org/10.1007/JHEP08(2016)043}{{\em JHEP} {\bfseries 08}
  (2016) 043},
\href{http://arxiv.org/abs/1602.06517}{{\ttfamily arXiv:1602.06517 [hep-th]}}.

\bibitem{Klaewer:2016kiy}
D.~Klaewer and E.~Palti, ``{Super-Planckian Spatial Field Variations and
  Quantum Gravity},'' \href{http://dx.doi.org/10.1007/JHEP01(2017)088}{{\em
  JHEP} {\bfseries 01} (2017) 088},
\href{http://arxiv.org/abs/1610.00010}{{\ttfamily arXiv:1610.00010 [hep-th]}}.

\bibitem{Valenzuela:2016yny}
I.~Valenzuela, ``{Backreaction Issues in Axion Monodromy and Minkowski
  4-forms},'' \href{http://dx.doi.org/10.1007/JHEP06(2017)098}{{\em JHEP}
  {\bfseries 06} (2017) 098},
\href{http://arxiv.org/abs/1611.00394}{{\ttfamily arXiv:1611.00394 [hep-th]}}.

\bibitem{Blumenhagen:2017cxt}
R.~Blumenhagen, I.~Valenzuela, and F.~Wolf, ``{The Swampland Conjecture and
  F-term Axion Monodromy Inflation},''
  \href{http://dx.doi.org/10.1007/JHEP07(2017)145}{{\em JHEP} {\bfseries 07}
  (2017) 145},
\href{http://arxiv.org/abs/1703.05776}{{\ttfamily arXiv:1703.05776 [hep-th]}}.

\bibitem{Palti:2017elp}
E.~Palti, ``{The Weak Gravity Conjecture and Scalar Fields},''
  \href{http://dx.doi.org/10.1007/JHEP08(2017)034}{{\em JHEP} {\bfseries 08}
  (2017) 034},
\href{http://arxiv.org/abs/1705.04328}{{\ttfamily arXiv:1705.04328 [hep-th]}}.

\bibitem{Lust:2017wrl}
{L\"{u}st, Dieter and Palti, Eran}, ``{Scalar Fields, Hierarchical UV/IR Mixing
  and The Weak Gravity Conjecture},''
  \href{http://dx.doi.org/10.1007/JHEP02(2018)040}{{\em JHEP} {\bfseries 02}
  (2018) 040},
\href{http://arxiv.org/abs/1709.01790}{{\ttfamily arXiv:1709.01790 [hep-th]}}.

\bibitem{Hebecker:2017lxm}
A.~Hebecker, P.~Henkenjohann, and L.~T. Witkowski, ``{Flat Monodromies and a
  Moduli Space Size Conjecture},''
  \href{http://dx.doi.org/10.1007/JHEP12(2017)033}{{\em JHEP} {\bfseries 12}
  (2017) 033},
\href{http://arxiv.org/abs/1708.06761}{{\ttfamily arXiv:1708.06761 [hep-th]}}.

\bibitem{Cicoli:2018tcq}
M.~Cicoli, D.~Ciupke, C.~Mayrhofer, and P.~Shukla, ``{A Geometrical Upper Bound
  on the Inflaton Range},''
  \href{http://dx.doi.org/10.1007/JHEP05(2018)001}{{\em JHEP} {\bfseries 05}
  (2018) 001},
\href{http://arxiv.org/abs/1801.05434}{{\ttfamily arXiv:1801.05434 [hep-th]}}.

\bibitem{Grimm:2018ohb}
T.~W. Grimm, E.~Palti, and I.~Valenzuela, ``{Infinite Distances in Field Space
  and Massless Towers of States},''
  \href{http://dx.doi.org/10.1007/JHEP08(2018)143}{{\em JHEP} {\bfseries 08}
  (2018) 143},
\href{http://arxiv.org/abs/1802.08264}{{\ttfamily arXiv:1802.08264 [hep-th]}}.

\bibitem{Heidenreich:2018kpg}
B.~Heidenreich, M.~Reece, and T.~Rudelius, ``{Emergence of Weak Coupling at
  Large Distance in Quantum Gravity},''
  \href{http://dx.doi.org/10.1103/PhysRevLett.121.051601}{{\em Phys. Rev.
  Lett.} {\bfseries 121} no.~5, (2018) 051601},
\href{http://arxiv.org/abs/1802.08698}{{\ttfamily arXiv:1802.08698 [hep-th]}}.

\bibitem{Blumenhagen:2018nts}
R.~Blumenhagen, D.~Klaewer, L.~Schlechter, and F.~Wolf, ``{The Refined
  Swampland Distance Conjecture in Calabi-Yau Moduli Spaces},''
  \href{http://dx.doi.org/10.1007/JHEP06(2018)052}{{\em JHEP} {\bfseries 06}
  (2018) 052},
\href{http://arxiv.org/abs/1803.04989}{{\ttfamily arXiv:1803.04989 [hep-th]}}.

\bibitem{Lee:2018urn}
S.-J. Lee, W.~Lerche, and T.~Weigand, ``{Tensionless Strings and the Weak
  Gravity Conjecture},''
\href{http://arxiv.org/abs/1808.05958}{{\ttfamily arXiv:1808.05958 [hep-th]}}.

\bibitem{Copeland:1994vg}
E.~J. Copeland, A.~R. Liddle, D.~H. Lyth, E.~D. Stewart, and D.~Wands, ``{False
  vacuum inflation with Einstein gravity},''
  \href{http://dx.doi.org/10.1103/PhysRevD.49.6410}{{\em Phys. Rev.} {\bfseries
  D49} (1994) 6410--6433},
\href{http://arxiv.org/abs/astro-ph/9401011}{{\ttfamily arXiv:astro-ph/9401011
  [astro-ph]}}.

\bibitem{Abazajian:2016yjj}
{\bfseries CMB-S4} Collaboration, K.~N. Abazajian {\em et~al.}, ``{CMB-S4
  Science Book, First Edition},''
\href{http://arxiv.org/abs/1610.02743}{{\ttfamily arXiv:1610.02743
  [astro-ph.CO]}}.

\bibitem{Suzuki:2018cuy}
A.~Suzuki {\em et~al.},
  \href{http://dx.doi.org/10.1007/s10909-018-1947-7}{``{The LiteBIRD Satellite
  Mission - Sub-Kelvin Instrument},''} in {\em {17th International Workshop on
  Low Temperature Detectors (LTD 17) Kurume City, Japan, July 17-21, 2017}}.
\newblock 2018.
\newblock
\href{http://arxiv.org/abs/1801.06987}{{\ttfamily arXiv:1801.06987
  [astro-ph.IM]}}.
\newblock

\bibitem{Kachru:2003sx}
S.~Kachru, R.~Kallosh, A.~D. Linde, J.~M. Maldacena, L.~P. McAllister, and
  S.~P. Trivedi, ``{Towards inflation in string theory},''
  \href{http://dx.doi.org/10.1088/1475-7516/2003/10/013}{{\em JCAP} {\bfseries
  0310} (2003) 013},
\href{http://arxiv.org/abs/hep-th/0308055}{{\ttfamily arXiv:hep-th/0308055
  [hep-th]}}.

\bibitem{Silverstein:2003hf}
E.~Silverstein and D.~Tong, ``{Scalar speed limits and cosmology: Acceleration
  from D-cceleration},''
  \href{http://dx.doi.org/10.1103/PhysRevD.70.103505}{{\em Phys. Rev.}
  {\bfseries D70} (2004) 103505},
\href{http://arxiv.org/abs/hep-th/0310221}{{\ttfamily arXiv:hep-th/0310221
  [hep-th]}}.

\bibitem{Balasubramanian:2005zx}
V.~Balasubramanian, P.~Berglund, J.~P. Conlon, and F.~Quevedo, ``{Systematics
  of moduli stabilisation in Calabi-Yau flux compactifications},''
  \href{http://dx.doi.org/10.1088/1126-6708/2005/03/007}{{\em JHEP} {\bfseries
  03} (2005) 007},
\href{http://arxiv.org/abs/hep-th/0502058}{{\ttfamily arXiv:hep-th/0502058
  [hep-th]}}.

\bibitem{Baumann:2006th}
D.~Baumann, A.~Dymarsky, I.~R. Klebanov, J.~M. Maldacena, L.~P. McAllister, and
  A.~Murugan, ``{On D3-brane Potentials in Compactifications with Fluxes and
  Wrapped D-branes},''
  \href{http://dx.doi.org/10.1088/1126-6708/2006/11/031}{{\em JHEP} {\bfseries
  11} (2006) 031},
\href{http://arxiv.org/abs/hep-th/0607050}{{\ttfamily arXiv:hep-th/0607050
  [hep-th]}}.

\bibitem{Westphal:2006tn}
A.~Westphal, ``{de Sitter string vacua from Kahler uplifting},''
  \href{http://dx.doi.org/10.1088/1126-6708/2007/03/102}{{\em JHEP} {\bfseries
  03} (2007) 102},
\href{http://arxiv.org/abs/hep-th/0611332}{{\ttfamily arXiv:hep-th/0611332
  [hep-th]}}.

\bibitem{Baumann:2007ah}
D.~Baumann, A.~Dymarsky, I.~R. Klebanov, and L.~McAllister, ``{Towards an
  Explicit Model of D-brane Inflation},''
  \href{http://dx.doi.org/10.1088/1475-7516/2008/01/024}{{\em JCAP} {\bfseries
  0801} (2008) 024},
\href{http://arxiv.org/abs/0706.0360}{{\ttfamily arXiv:0706.0360 [hep-th]}}.

\bibitem{Dong:2010pm}
X.~Dong, B.~Horn, E.~Silverstein, and G.~Torroba, ``{Micromanaging de Sitter
  holography},'' \href{http://dx.doi.org/10.1088/0264-9381/27/24/245020}{{\em
  Class. Quant. Grav.} {\bfseries 27} (2010) 245020},
\href{http://arxiv.org/abs/1005.5403}{{\ttfamily arXiv:1005.5403 [hep-th]}}.

\bibitem{Rummel:2011cd}
M.~Rummel and A.~Westphal, ``{A sufficient condition for de Sitter vacua in
  type IIB string theory},''
  \href{http://dx.doi.org/10.1007/JHEP01(2012)020}{{\em JHEP} {\bfseries 01}
  (2012) 020},
\href{http://arxiv.org/abs/1107.2115}{{\ttfamily arXiv:1107.2115 [hep-th]}}.

\bibitem{Blaback:2013fca}
J.~Bl\r{a}b{\"{a}}ck, U.~Danielsson, and G.~Dibitetto, ``{Accelerated Universes
  from type IIA Compactifications},''
  \href{http://dx.doi.org/10.1088/1475-7516/2014/03/003}{{\em JCAP} {\bfseries
  1403} (2014) 003},
\href{http://arxiv.org/abs/1310.8300}{{\ttfamily arXiv:1310.8300 [hep-th]}}.

\bibitem{Cicoli:2013cha}
M.~Cicoli, D.~Klevers, S.~Krippendorf, C.~Mayrhofer, F.~Quevedo, and
  R.~Valandro, ``{Explicit de Sitter Flux Vacua for Global String Models with
  Chiral Matter},'' \href{http://dx.doi.org/10.1007/JHEP05(2014)001}{{\em JHEP}
  {\bfseries 05} (2014) 001},
\href{http://arxiv.org/abs/1312.0014}{{\ttfamily arXiv:1312.0014 [hep-th]}}.

\bibitem{Cicoli:2015ylx}
M.~Cicoli, F.~Quevedo, and R.~Valandro, ``{De Sitter from T-branes},''
  \href{http://dx.doi.org/10.1007/JHEP03(2016)141}{{\em JHEP} {\bfseries 03}
  (2016) 141},
\href{http://arxiv.org/abs/1512.04558}{{\ttfamily arXiv:1512.04558 [hep-th]}}.

\bibitem{Maldacena:2000mw}
J.~M. Maldacena and C.~Nunez, ``{Supergravity description of field theories on
  curved manifolds and a no go theorem},''
  \href{http://dx.doi.org/10.1142/S0217751X01003935,
  10.1142/S0217751X01003937}{{\em Int. J. Mod. Phys.} {\bfseries A16} (2001)
  822--855}, \href{http://arxiv.org/abs/hep-th/0007018}{{\ttfamily
  arXiv:hep-th/0007018 [hep-th]}}.
[,182(2000)].

\bibitem{Townsend:2003qv}
P.~K. Townsend, ``{Cosmic acceleration and M theory},'' in {\em {Mathematical
  physics. Proceedings, 14th International Congress, ICMP 2003, Lisbon,
  Portugal, July 28-August 2, 2003}}, pp.~655--662.
\newblock 2003.
\newblock
\href{http://arxiv.org/abs/hep-th/0308149}{{\ttfamily arXiv:hep-th/0308149
  [hep-th]}}.
\newblock

\bibitem{Hertzberg:2007wc}
M.~P. Hertzberg, S.~Kachru, W.~Taylor, and M.~Tegmark, ``{Inflationary
  Constraints on Type IIA String Theory},''
  \href{http://dx.doi.org/10.1088/1126-6708/2007/12/095}{{\em JHEP} {\bfseries
  12} (2007) 095},
\href{http://arxiv.org/abs/0711.2512}{{\ttfamily arXiv:0711.2512 [hep-th]}}.

\bibitem{Covi:2008ea}
L.~Covi, M.~Gomez-Reino, C.~Gross, J.~Louis, G.~A. Palma, and C.~A. Scrucca,
  ``{de Sitter vacua in no-scale supergravities and Calabi-Yau string
  models},'' \href{http://dx.doi.org/10.1088/1126-6708/2008/06/057}{{\em JHEP}
  {\bfseries 06} (2008) 057},
\href{http://arxiv.org/abs/0804.1073}{{\ttfamily arXiv:0804.1073 [hep-th]}}.

\bibitem{McGuirk:2009xx}
P.~McGuirk, G.~Shiu, and Y.~Sumitomo, ``{Non-supersymmetric infrared
  perturbations to the warped deformed conifold},''
  \href{http://dx.doi.org/10.1016/j.nuclphysb.2010.09.008}{{\em Nucl. Phys.}
  {\bfseries B842} (2011) 383--413},
\href{http://arxiv.org/abs/0910.4581}{{\ttfamily arXiv:0910.4581 [hep-th]}}.

\bibitem{Caviezel:2008tf}
{Caviezel, Claudio and Koerber, Paul and Kors, Simon and L\"{u}st, Dieter and
  Wrase, Timm and Zagermann, Marco}, ``{On the Cosmology of Type IIA
  Compactifications on SU(3)-structure Manifolds},''
  \href{http://dx.doi.org/10.1088/1126-6708/2009/04/010}{{\em JHEP} {\bfseries
  04} (2009) 010},
\href{http://arxiv.org/abs/0812.3551}{{\ttfamily arXiv:0812.3551 [hep-th]}}.

\bibitem{Caviezel:2009tu}
C.~Caviezel, T.~Wrase, and M.~Zagermann, ``{Moduli Stabilization and Cosmology
  of Type IIB on SU(2)-Structure Orientifolds},''
  \href{http://dx.doi.org/10.1007/JHEP04(2010)011}{{\em JHEP} {\bfseries 04}
  (2010) 011},
\href{http://arxiv.org/abs/0912.3287}{{\ttfamily arXiv:0912.3287 [hep-th]}}.

\bibitem{deCarlos:2009fq}
B.~de~Carlos, A.~Guarino, and J.~M. Moreno, ``{Flux moduli stabilisation,
  Supergravity algebras and no-go theorems},''
  \href{http://dx.doi.org/10.1007/JHEP01(2010)012}{{\em JHEP} {\bfseries 01}
  (2010) 012},
\href{http://arxiv.org/abs/0907.5580}{{\ttfamily arXiv:0907.5580 [hep-th]}}.

\bibitem{Wrase:2010ew}
T.~Wrase and M.~Zagermann, ``{On Classical de Sitter Vacua in String Theory},''
  \href{http://dx.doi.org/10.1002/prop.201000053}{{\em Fortsch. Phys.}
  {\bfseries 58} (2010) 906--910},
\href{http://arxiv.org/abs/1003.0029}{{\ttfamily arXiv:1003.0029 [hep-th]}}.

\bibitem{Shiu:2011zt}
G.~Shiu and Y.~Sumitomo, ``{Stability Constraints on Classical de Sitter
  Vacua},'' \href{http://dx.doi.org/10.1007/JHEP09(2011)052}{{\em JHEP}
  {\bfseries 09} (2011) 052},
\href{http://arxiv.org/abs/1107.2925}{{\ttfamily arXiv:1107.2925 [hep-th]}}.

\bibitem{Green:2011cn}
S.~R. Green, E.~J. Martinec, C.~Quigley, and S.~Sethi, ``{Constraints on String
  Cosmology},'' \href{http://dx.doi.org/10.1088/0264-9381/29/7/075006}{{\em
  Class. Quant. Grav.} {\bfseries 29} (2012) 075006},
\href{http://arxiv.org/abs/1110.0545}{{\ttfamily arXiv:1110.0545 [hep-th]}}.

\bibitem{Gautason:2012tb}
F.~F. Gautason, D.~Junghans, and M.~Zagermann, ``{On Cosmological Constants
  from alpha'-Corrections},''
  \href{http://dx.doi.org/10.1007/JHEP06(2012)029}{{\em JHEP} {\bfseries 06}
  (2012) 029},
\href{http://arxiv.org/abs/1204.0807}{{\ttfamily arXiv:1204.0807 [hep-th]}}.

\bibitem{Bena:2012vz}
I.~Bena, M.~Grana, S.~Kuperstein, and S.~Massai, ``{Polchinski-Strassler does
  not uplift Klebanov-Strassler},''
  \href{http://dx.doi.org/10.1007/JHEP09(2013)142}{{\em JHEP} {\bfseries 09}
  (2013) 142},
\href{http://arxiv.org/abs/1212.4828}{{\ttfamily arXiv:1212.4828 [hep-th]}}.

\bibitem{Blaback:2012nf}
J.~Bl\r{a}b{\"{a}}ck, U.~H. Danielsson, and T.~Van~Riet, ``{Resolving
  anti-brane singularities through time-dependence},''
  \href{http://dx.doi.org/10.1007/JHEP02(2013)061}{{\em JHEP} {\bfseries 02}
  (2013) 061},
\href{http://arxiv.org/abs/1202.1132}{{\ttfamily arXiv:1202.1132 [hep-th]}}.

\bibitem{Bena:2014jaa}
I.~Bena, M.~Grana, S.~Kuperstein, and S.~Massai, ``{Giant Tachyons in the
  Landscape},'' \href{http://dx.doi.org/10.1007/JHEP02(2015)146}{{\em JHEP}
  {\bfseries 02} (2015) 146},
\href{http://arxiv.org/abs/1410.7776}{{\ttfamily arXiv:1410.7776 [hep-th]}}.

\bibitem{Danielsson:2014yga}
U.~H. Danielsson and T.~Van~Riet, ``{Fatal attraction: more on decaying
  anti-branes},'' \href{http://dx.doi.org/10.1007/JHEP03(2015)087}{{\em JHEP}
  {\bfseries 03} (2015) 087},
\href{http://arxiv.org/abs/1410.8476}{{\ttfamily arXiv:1410.8476 [hep-th]}}.

\bibitem{Kutasov:2015eba}
D.~Kutasov, T.~Maxfield, I.~Melnikov, and S.~Sethi, ``{Constraining de Sitter
  Space in String Theory},''
  \href{http://dx.doi.org/10.1103/PhysRevLett.115.071305}{{\em Phys. Rev.
  Lett.} {\bfseries 115} no.~7, (2015) 071305},
\href{http://arxiv.org/abs/1504.00056}{{\ttfamily arXiv:1504.00056 [hep-th]}}.

\bibitem{Quigley:2015jia}
C.~Quigley, ``{Gaugino Condensation and the Cosmological Constant},''
  \href{http://dx.doi.org/10.1007/JHEP06(2015)104}{{\em JHEP} {\bfseries 06}
  (2015) 104},
\href{http://arxiv.org/abs/1504.00652}{{\ttfamily arXiv:1504.00652 [hep-th]}}.

\bibitem{Dasgupta:2014pma}
K.~Dasgupta, R.~Gwyn, E.~McDonough, M.~Mia, and R.~Tatar, ``{de Sitter Vacua in
  Type IIB String Theory: Classical Solutions and Quantum Corrections},''
  \href{http://dx.doi.org/10.1007/JHEP07(2014)054}{{\em JHEP} {\bfseries 07}
  (2014) 054},
\href{http://arxiv.org/abs/1402.5112}{{\ttfamily arXiv:1402.5112 [hep-th]}}.

\bibitem{Junghans:2016abx}
D.~Junghans and M.~Zagermann, ``{A Universal Tachyon in Nearly No-scale de
  Sitter Compactifications},''
  \href{http://dx.doi.org/10.1007/JHEP07(2018)078}{{\em JHEP} {\bfseries 07}
  (2018) 078},
\href{http://arxiv.org/abs/1612.06847}{{\ttfamily arXiv:1612.06847 [hep-th]}}.

\bibitem{Junghans:2016uvg}
D.~Junghans, ``{Tachyons in Classical de Sitter Vacua},''
  \href{http://dx.doi.org/10.1007/JHEP06(2016)132}{{\em JHEP} {\bfseries 06}
  (2016) 132},
\href{http://arxiv.org/abs/1603.08939}{{\ttfamily arXiv:1603.08939 [hep-th]}}.

\bibitem{Andriot:2016xvq}
D.~Andriot and J.~Bl\r{a}b{\"{a}}ck, ``{Refining the boundaries of the
  classical de Sitter landscape},''
  \href{http://dx.doi.org/10.1007/JHEP03(2017)102,
  10.1007/JHEP03(2018)083}{{\em JHEP} {\bfseries 03} (2017) 102},
  \href{http://arxiv.org/abs/1609.00385}{{\ttfamily arXiv:1609.00385
  [hep-th]}}.
[Erratum: JHEP03,083(2018)].

\bibitem{Moritz:2017xto}
J.~Moritz, A.~Retolaza, and A.~Westphal, ``{Toward de Sitter space from ten
  dimensions},'' \href{http://dx.doi.org/10.1103/PhysRevD.97.046010}{{\em Phys.
  Rev.} {\bfseries D97} no.~4, (2018) 046010},
\href{http://arxiv.org/abs/1707.08678}{{\ttfamily arXiv:1707.08678 [hep-th]}}.

\bibitem{Sethi:2017phn}
S.~Sethi, ``{Supersymmetry Breaking by Fluxes},''
  \href{http://dx.doi.org/10.1007/JHEP10(2018)022}{{\em JHEP} {\bfseries 10}
  (2018) 022},
\href{http://arxiv.org/abs/1709.03554}{{\ttfamily arXiv:1709.03554 [hep-th]}}.

\bibitem{Andriot:2017jhf}
D.~Andriot, ``{On classical de Sitter and Minkowski solutions with intersecting
  branes},'' \href{http://dx.doi.org/10.1007/JHEP03(2018)054}{{\em JHEP}
  {\bfseries 03} (2018) 054},
\href{http://arxiv.org/abs/1710.08886}{{\ttfamily arXiv:1710.08886 [hep-th]}}.

\bibitem{Danielsson:2018ztv}
U.~H. Danielsson and T.~Van~Riet, ``{What if string theory has no de Sitter
  vacua?},'' \href{http://dx.doi.org/10.1142/S0218271818300070}{{\em Int. J.
  Mod. Phys.} {\bfseries D27} no.~12, (2018) 1830007},
\href{http://arxiv.org/abs/1804.01120}{{\ttfamily arXiv:1804.01120 [hep-th]}}.

\bibitem{Obied:2018sgi}
G.~Obied, H.~Ooguri, L.~Spodyneiko, and C.~Vafa, ``{De Sitter Space and the
  Swampland},''
\href{http://arxiv.org/abs/1806.08362}{{\ttfamily arXiv:1806.08362 [hep-th]}}.

\bibitem{Kallosh:2014wsa}
R.~Kallosh and T.~Wrase, ``{Emergence of Spontaneously Broken Supersymmetry on
  an Anti-D3-Brane in KKLT dS Vacua},''
  \href{http://dx.doi.org/10.1007/JHEP12(2014)117}{{\em JHEP} {\bfseries 12}
  (2014) 117},
\href{http://arxiv.org/abs/1411.1121}{{\ttfamily arXiv:1411.1121 [hep-th]}}.

\bibitem{Ferrara:2014kva}
S.~Ferrara, R.~Kallosh, and A.~Linde, ``{Cosmology with Nilpotent
  Superfields},'' \href{http://dx.doi.org/10.1007/JHEP10(2014)143}{{\em JHEP}
  {\bfseries 10} (2014) 143},
\href{http://arxiv.org/abs/1408.4096}{{\ttfamily arXiv:1408.4096 [hep-th]}}.

\bibitem{Bergshoeff:2015jxa}
E.~A. Bergshoeff, K.~Dasgupta, R.~Kallosh, A.~Van~Proeyen, and T.~Wrase, ``{$
  \overline{\mathrm{D}3} $ and dS},''
  \href{http://dx.doi.org/10.1007/JHEP05(2015)058}{{\em JHEP} {\bfseries 05}
  (2015) 058},
\href{http://arxiv.org/abs/1502.07627}{{\ttfamily arXiv:1502.07627 [hep-th]}}.

\bibitem{Bergshoeff:2015tra}
E.~A. Bergshoeff, D.~Z. Freedman, R.~Kallosh, and A.~Van~Proeyen, ``{Pure de
  Sitter Supergravity},'' \href{http://dx.doi.org/10.1103/PhysRevD.93.069901,
  10.1103/PhysRevD.92.085040}{{\em Phys. Rev.} {\bfseries D92} no.~8, (2015)
  085040}, \href{http://arxiv.org/abs/1507.08264}{{\ttfamily arXiv:1507.08264
  [hep-th]}}.
[Erratum: Phys. Rev.D93,no.6,069901(2016)].

\bibitem{Hasegawa:2015bza}
F.~Hasegawa and Y.~Yamada, ``{Component action of nilpotent multiplet coupled
  to matter in 4 dimensional $ \mathcal{N}=1 $ supergravity},''
  \href{http://dx.doi.org/10.1007/JHEP10(2015)106}{{\em JHEP} {\bfseries 10}
  (2015) 106},
\href{http://arxiv.org/abs/1507.08619}{{\ttfamily arXiv:1507.08619 [hep-th]}}.

\bibitem{Kallosh:2015nia}
R.~Kallosh, F.~Quevedo, and A.~M. Uranga, ``{String Theory Realizations of the
  Nilpotent Goldstino},'' \href{http://dx.doi.org/10.1007/JHEP12(2015)039}{{\em
  JHEP} {\bfseries 12} (2015) 039},
\href{http://arxiv.org/abs/1507.07556}{{\ttfamily arXiv:1507.07556 [hep-th]}}.

\bibitem{Polchinski:2015bea}
J.~Polchinski, ``{Brane/antibrane dynamics and KKLT stability},''
\href{http://arxiv.org/abs/1509.05710}{{\ttfamily arXiv:1509.05710 [hep-th]}}.

\bibitem{Kallosh:2016aep}
R.~Kallosh, B.~Vercnocke, and T.~Wrase, ``{String Theory Origin of Constrained
  Multiplets},'' \href{http://dx.doi.org/10.1007/JHEP09(2016)063}{{\em JHEP}
  {\bfseries 09} (2016) 063},
\href{http://arxiv.org/abs/1606.09245}{{\ttfamily arXiv:1606.09245 [hep-th]}}.

\bibitem{Aalsma:2018pll}
L.~Aalsma, M.~Tournoy, J.~P. Van Der~Schaar, and B.~Vercnocke, ``{A
  Supersymmetric Embedding of Anti-Brane Polarization},''
\href{http://arxiv.org/abs/1807.03303}{{\ttfamily arXiv:1807.03303 [hep-th]}}.

\bibitem{Cicoli:2018kdo}
M.~Cicoli, S.~de~Alwis, A.~Maharana, F.~Muia, and F.~Quevedo, ``{De Sitter vs
  Quintessence in String Theory},''
\href{http://arxiv.org/abs/1808.08967}{{\ttfamily arXiv:1808.08967 [hep-th]}}.

\bibitem{Dasgupta:2018rtp}
K.~Dasgupta, M.~Emelin, E.~McDonough, and R.~Tatar, ``{Quantum Corrections and
  the de Sitter Swampland Conjecture},''
\href{http://arxiv.org/abs/1808.07498}{{\ttfamily arXiv:1808.07498 [hep-th]}}.

\bibitem{Roupec:2018mbn}
C.~Roupec and T.~Wrase, ``{de Sitter extrema and the swampland},''
\href{http://arxiv.org/abs/1807.09538}{{\ttfamily arXiv:1807.09538 [hep-th]}}.

\bibitem{Andriot:2018ept}
D.~Andriot, ``{New constraints on classical de Sitter: flirting with the
  swampland},''
\href{http://arxiv.org/abs/1807.09698}{{\ttfamily arXiv:1807.09698 [hep-th]}}.

\bibitem{Andriot:2018wzk}
D.~Andriot, ``{On the de Sitter swampland criterion},''
\href{http://arxiv.org/abs/1806.10999}{{\ttfamily arXiv:1806.10999 [hep-th]}}.

\bibitem{Conlon:2018eyr}
J.~P. Conlon, ``{The de Sitter swampland conjecture and supersymmetric AdS
  vacua},''
\href{http://arxiv.org/abs/1808.05040}{{\ttfamily arXiv:1808.05040 [hep-th]}}.

\bibitem{Akrami:2018ylq}
Y.~Akrami, R.~Kallosh, A.~Linde, and V.~Vardanyan, ``{The landscape, the
  swampland and the era of precision cosmology},''
\href{http://arxiv.org/abs/1808.09440}{{\ttfamily arXiv:1808.09440 [hep-th]}}.

\bibitem{Kallosh:2018wme}
R.~Kallosh, A.~Linde, E.~McDonough, and M.~Scalisi, ``{de Sitter Vacua with a
  Nilpotent Superfield},''
\href{http://arxiv.org/abs/1808.09428}{{\ttfamily arXiv:1808.09428 [hep-th]}}.

\bibitem{Kachru:2018aqn}
S.~Kachru and S.~Trivedi, ``{A comment on effective field theories of flux
  vacua},''
\href{http://arxiv.org/abs/1808.08971}{{\ttfamily arXiv:1808.08971 [hep-th]}}.

\bibitem{Moritz:2018ani}
J.~Moritz, A.~Retolaza, and A.~Westphal, ``{On uplifts by warped
  anti-D3-branes},''
\href{http://arxiv.org/abs/1809.06618}{{\ttfamily arXiv:1809.06618 [hep-th]}}.

\bibitem{Bena:2018fqc}
I.~Bena, E.~Dudas, M.~Gra{\~n}a, and S.~L{\"u}st, ``{Uplifting Runaways},''
\href{http://arxiv.org/abs/1809.06861}{{\ttfamily arXiv:1809.06861 [hep-th]}}.

\bibitem{Gautason:2018gln}
F.~F. Gautason, V.~Van~Hemelryck, and T.~Van~Riet, ``{The tension between 10D
  supergravity and dS uplifts},''
\href{http://arxiv.org/abs/1810.08518}{{\ttfamily arXiv:1810.08518 [hep-th]}}.

\bibitem{Choi:2018rze}
K.~Choi, D.~Chway, and C.~S. Shin, ``{The dS swampland conjecture with the
  electroweak symmetry and QCD chiral symmetry breaking},''
\href{http://arxiv.org/abs/1809.01475}{{\ttfamily arXiv:1809.01475 [hep-th]}}.

\bibitem{Heisenberg:2018rdu}
L.~Heisenberg, M.~Bartelmann, R.~Brandenberger, and A.~Refregier, ``{Dark
  Energy in the Swampland II},''
\href{http://arxiv.org/abs/1809.00154}{{\ttfamily arXiv:1809.00154
  [astro-ph.CO]}}.

\bibitem{Heisenberg:2018yae}
L.~Heisenberg, M.~Bartelmann, R.~Brandenberger, and A.~Refregier, ``{Dark
  Energy in the Swampland},''
\href{http://arxiv.org/abs/1808.02877}{{\ttfamily arXiv:1808.02877
  [astro-ph.CO]}}.

\bibitem{Marsh:2018kub}
M.~C.~D. Marsh, ``{The Swampland, Quintessence and the Vacuum Energy},''
\href{http://arxiv.org/abs/1809.00726}{{\ttfamily arXiv:1809.00726 [hep-th]}}.

\bibitem{Murayama:2018lie}
H.~Murayama, M.~Yamazaki, and T.~T. Yanagida, ``{Do We Live in the
  Swampland?},''
\href{http://arxiv.org/abs/1809.00478}{{\ttfamily arXiv:1809.00478 [hep-th]}}.

\bibitem{Kinney:2018nny}
W.~H. Kinney, S.~Vagnozzi, and L.~Visinelli, ``{The Zoo Plot Meets the
  Swampland: Mutual (In)Consistency of Single-Field Inflation, String
  Conjectures, and Cosmological Data},''
\href{http://arxiv.org/abs/1808.06424}{{\ttfamily arXiv:1808.06424
  [astro-ph.CO]}}.

\bibitem{Damian:2018tlf}
O.~Loaiza-Brito and O.~Loaiza-Brito, ``{Two-field axion inflation and the
  swampland constraint in the flux-scaling scenario},''
\href{http://arxiv.org/abs/1808.03397}{{\ttfamily arXiv:1808.03397 [hep-th]}}.

\bibitem{Ben-Dayan:2018mhe}
I.~Ben-Dayan, ``{Draining the Swampland},''
\href{http://arxiv.org/abs/1808.01615}{{\ttfamily arXiv:1808.01615 [hep-th]}}.

\bibitem{Matsui:2018bsy}
H.~Matsui and F.~Takahashi, ``{Eternal Inflation and Swampland Conjectures},''
\href{http://arxiv.org/abs/1807.11938}{{\ttfamily arXiv:1807.11938 [hep-th]}}.

\bibitem{Colgain:2018wgk}
O.~E. Colg\'a~in, M.~H. P.~M. Van~Putten, and H.~Yavartanoo, ``{$H_0$ tension
  and the de Sitter Swampland},''
\href{http://arxiv.org/abs/1807.07451}{{\ttfamily arXiv:1807.07451 [hep-th]}}.

\bibitem{Denef:2018etk}
F.~Denef, A.~Hebecker, and T.~Wrase, ``{The dS swampland conjecture and the
  Higgs potential},''
\href{http://arxiv.org/abs/1807.06581}{{\ttfamily arXiv:1807.06581 [hep-th]}}.

\bibitem{Dias:2018ngv}
M.~Dias, J.~Frazer, A.~Retolaza, and A.~Westphal, ``{Primordial Gravitational
  Waves and the Swampland},''
\href{http://arxiv.org/abs/1807.06579}{{\ttfamily arXiv:1807.06579 [hep-th]}}.

\bibitem{Kehagias:2018uem}
A.~Kehagias and A.~Riotto, ``{A note on Inflation and the Swampland},''
\href{http://arxiv.org/abs/1807.05445}{{\ttfamily arXiv:1807.05445 [hep-th]}}.

\bibitem{Garg:2018reu}
S.~K. Garg and C.~Krishnan, ``{Bounds on Slow Roll and the de Sitter
  Swampland},''
\href{http://arxiv.org/abs/1807.05193}{{\ttfamily arXiv:1807.05193 [hep-th]}}.

\bibitem{Achucarro:2018vey}
A.~Ach\'ucarro and G.~A. Palma, ``{The string swampland constraints require
  multi-field inflation},''
\href{http://arxiv.org/abs/1807.04390}{{\ttfamily arXiv:1807.04390 [hep-th]}}.

\bibitem{Das:2018hqy}
S.~Das, ``{A note on Single-field Inflation and the Swampland Criteria},''
\href{http://arxiv.org/abs/1809.03962}{{\ttfamily arXiv:1809.03962 [hep-th]}}.

\bibitem{Wang:2018duq}
D.~Wang, ``{The multi-feature universe: large parameter space cosmology and the
  swampland},''
\href{http://arxiv.org/abs/1809.04854}{{\ttfamily arXiv:1809.04854
  [astro-ph.CO]}}.

\bibitem{Brandenberger:2018wbg}
R.~H. Brandenberger, ``{Beyond Standard Inflationary Cosmology},''
\href{http://arxiv.org/abs/1809.04926}{{\ttfamily arXiv:1809.04926 [hep-th]}}.

\bibitem{Han:2018yrk}
C.~Han, S.~Pi, and M.~Sasaki, ``{Quintessence Saves Higgs Instability},''
\href{http://arxiv.org/abs/1809.05507}{{\ttfamily arXiv:1809.05507 [hep-ph]}}.

\bibitem{Brandenberger:2018xnf}
R.~Brandenberger, R.~R. Cuzinatto, J.~Fr{\"o}hlich, and R.~Namba, ``{New Scalar
  Field Quartessence},''
\href{http://arxiv.org/abs/1809.07409}{{\ttfamily arXiv:1809.07409 [gr-qc]}}.

\bibitem{Dimopoulos:2018upl}
K.~Dimopoulos, ``{Steep Eternal Inflation and the Swampland},''
\href{http://arxiv.org/abs/1810.03438}{{\ttfamily arXiv:1810.03438 [gr-qc]}}.

\bibitem{Ellis:2018xdr}
J.~Ellis, B.~Nagaraj, D.~V. Nanopoulos, and K.~A. Olive, ``{De Sitter Vacua in
  No-Scale Supergravity},''
\href{http://arxiv.org/abs/1809.10114}{{\ttfamily arXiv:1809.10114 [hep-th]}}.

\bibitem{Lin:2018kjm}
C.-M. Lin, K.-W. Ng, and K.~Cheung, ``{Chaotic inflation on the brane and the
  Swampland Criteria},''
\href{http://arxiv.org/abs/1810.01644}{{\ttfamily arXiv:1810.01644 [hep-ph]}}.

\bibitem{Hamaguchi:2018vtv}
K.~Hamaguchi, M.~Ibe, and T.~Moroi, ``{The swampland conjecture and the Higgs
  expectation value},''
\href{http://arxiv.org/abs/1810.02095}{{\ttfamily arXiv:1810.02095 [hep-th]}}.

\bibitem{Kawasaki:2018daf}
M.~Kawasaki and V.~Takhistov, ``{Primordial Black Holes and the String
  Swampland},''
\href{http://arxiv.org/abs/1810.02547}{{\ttfamily arXiv:1810.02547 [hep-th]}}.

\bibitem{Motaharfar:2018zyb}
M.~Motaharfar, V.~Kamali, and R.~O. Ramos, ``{Warm way out of the Swampland},''
\href{http://arxiv.org/abs/1810.02816}{{\ttfamily arXiv:1810.02816
  [astro-ph.CO]}}.

\bibitem{Ashoorioon:2018sqb}
A.~Ashoorioon, ``{Rescuing Single Field Inflation from the Swampland},''
\href{http://arxiv.org/abs/1810.04001}{{\ttfamily arXiv:1810.04001 [hep-th]}}.

\bibitem{Das:2018rpg}
S.~Das, ``{Warm Inflation in the light of Swampland Criteria},''
\href{http://arxiv.org/abs/1810.05038}{{\ttfamily arXiv:1810.05038 [hep-th]}}.

\bibitem{Wang:2018kly}
S.-J. Wang, ``{Quintessential Starobinsky inflation and swampland criteria},''
\href{http://arxiv.org/abs/1810.06445}{{\ttfamily arXiv:1810.06445 [hep-th]}}.

\bibitem{Fukuda:2018haz}
H.~Fukuda, R.~Saito, S.~Shirai, and M.~Yamazaki, ``{Phenomenological
  Consequences of the Refined Swampland Conjecture},''
\href{http://arxiv.org/abs/1810.06532}{{\ttfamily arXiv:1810.06532 [hep-th]}}.

\bibitem{Hebecker:2018vxz}
A.~Hebecker and T.~Wrase, ``{The asymptotic dS Swampland Conjecture - a
  simplified derivation and a potential loophole},''
\href{http://arxiv.org/abs/1810.08182}{{\ttfamily arXiv:1810.08182 [hep-th]}}.

\bibitem{Olguin-Tejo:2018pfq}
Y.~Olguin-Tejo, S.~L. Parameswaran, G.~Tasinato, and I.~Zavala, ``{Runaway
  Quintessence, Out of the Swampland},''
\href{http://arxiv.org/abs/1810.08634}{{\ttfamily arXiv:1810.08634 [hep-th]}}.

\bibitem{Garg:2018zdg}
S.~K. Garg, C.~Krishnan, and M.~Z. Zaz, ``{Bounds on Slow Roll at the Boundary
  of the Landscape},''
\href{http://arxiv.org/abs/1810.09406}{{\ttfamily arXiv:1810.09406 [hep-th]}}.

\bibitem{Park:2018fuj}
S.~C. Park, ``{Minimal gauge inflation and the refined Swampland conjecture},''
\href{http://arxiv.org/abs/1810.11279}{{\ttfamily arXiv:1810.11279 [hep-ph]}}.

\bibitem{Blaback:2018hdo}
J.~Bl\r{a}b{\"{a}}ck, U.~Danielsson, and G.~Dibitetto, ``{A new light on the
  darkest corner of the landscape},''
\href{http://arxiv.org/abs/1810.11365}{{\ttfamily arXiv:1810.11365 [hep-th]}}.

\bibitem{Schimmrigk:2018gch}
R.~Schimmrigk, ``{The Swampland Spectrum Conjecture in Inflation},''
\href{http://arxiv.org/abs/1810.11699}{{\ttfamily arXiv:1810.11699 [hep-th]}}.

\bibitem{Lin:2018rnx}
C.-M. Lin, ``{Type I Hilltop Inflation and the Refined Swampland Criteria},''
\href{http://arxiv.org/abs/1810.11992}{{\ttfamily arXiv:1810.11992
  [astro-ph.CO]}}.

\bibitem{Agrawal:2018own}
P.~Agrawal, G.~Obied, P.~J. Steinhardt, and C.~Vafa, ``{On the Cosmological
  Implications of the String Swampland},''
\href{http://arxiv.org/abs/1806.09718}{{\ttfamily arXiv:1806.09718 [hep-th]}}.

\bibitem{Brax:2009kd}
P.~Brax, C.~van~de Bruck, J.~Martin, and A.-C. Davis, ``{Decoupling Dark Energy
  from Matter},'' \href{http://dx.doi.org/10.1088/1475-7516/2009/09/032}{{\em
  JCAP} {\bfseries 0909} (2009) 032},
\href{http://arxiv.org/abs/0904.3471}{{\ttfamily arXiv:0904.3471 [hep-th]}}.

\bibitem{Chiang:2018jdg}
C.-I. Chiang and H.~Murayama, ``{Building Supergravity Quintessence Model},''
\href{http://arxiv.org/abs/1808.02279}{{\ttfamily arXiv:1808.02279 [hep-th]}}.

\bibitem{Ooguri:2018wrx}
H.~Ooguri, E.~Palti, G.~Shiu, and C.~Vafa, ``{Distance and de Sitter
  Conjectures on the Swampland},''
\href{http://arxiv.org/abs/1810.05506}{{\ttfamily arXiv:1810.05506 [hep-th]}}.

\bibitem{Agrawal:2018rcg}
P.~Agrawal and G.~Obied, ``{Dark Energy and the Refined de Sitter
  Conjecture},''
\href{http://arxiv.org/abs/1811.00554}{{\ttfamily arXiv:1811.00554 [hep-ph]}}.

\bibitem{Akrami:2018odb}
{\bfseries Planck} Collaboration, Y.~Akrami {\em et~al.}, ``{Planck 2018
  results. X. Constraints on inflation},''
\href{http://arxiv.org/abs/1807.06211}{{\ttfamily arXiv:1807.06211
  [astro-ph.CO]}}.

\bibitem{Hetz:2016ics}
A.~Hetz and G.~A. Palma, ``{Sound Speed of Primordial Fluctuations in
  Supergravity Inflation},''
  \href{http://dx.doi.org/10.1103/PhysRevLett.117.101301}{{\em Phys. Rev.
  Lett.} {\bfseries 117} no.~10, (2016) 101301},
\href{http://arxiv.org/abs/1601.05457}{{\ttfamily arXiv:1601.05457 [hep-th]}}.

\bibitem{Ade:2015ava}
{\bfseries Planck} Collaboration, P.~A.~R. Ade {\em et~al.}, ``{Planck 2015
  results. XVII. Constraints on primordial non-Gaussianity},''
  \href{http://dx.doi.org/10.1051/0004-6361/201525836}{{\em Astron. Astrophys.}
  {\bfseries 594} (2016) A17},
\href{http://arxiv.org/abs/1502.01592}{{\ttfamily arXiv:1502.01592
  [astro-ph.CO]}}.

\bibitem{CMBS4test}
{\bfseries Cosmic Microwave Background Stage 4 Concept Definition Task Force}
  Collaboration, C.~Lawrence {\em et~al.}, ``{Report To The Astronomy and
  Astrophysics Advisory Committee}.''
  \url{https://www.nsf.gov/mps/ast/aaac/cmb_s4/report/CMBS4_final_report_NL.pdf}.

\bibitem{DES}
{\bfseries {Dark Energy Survey}} Collaboration.
  \url{https://www.darkenergysurvey.org/}.

\bibitem{HSC}
{\bfseries {Hyper Suprime-Cam}} Collaboration.
  \url{https://hsc.mtk.nao.ac.jp/ssp/}.

\bibitem{DESI}
{\bfseries {Dark Energy Spectroscopic Instrument}} Collaboration.
  \url{https://www.desi.lbl.gov}.

\bibitem{PFS}
{\bfseries {Prime Focus Spectrograph}} Collaboration.
  \url{https://pfs.ipmu.jp}.

\bibitem{LSST}
{\bfseries {Large Synoptic Survey Telescope}} Collaboration.
  \url{https://www.lsstcorporation.org}.

\bibitem{Euclid}
{\bfseries {Euclid}} Collaboration. \url{https://www.euclid-ec.org}.

\bibitem{WFIRST}
{\bfseries {Wide Field Infrared Survey Telescope}} Collaboration.
  \url{https://wfirst.gsfc.nasa.gov}.

\end{thebibliography}\endgroup

\end{document}